\documentclass[twocolumn,pra,showkeys,aps,showpacs,10pt]{revtex4-1}
\usepackage{amssymb}
\usepackage{graphicx}
\usepackage{natbib} 
\usepackage{mathptmx}
\begin{document}
\title {Scaling Conception of Energy Loss' Separation in Soft Magnetic Materials}  
\author{Krzysztof \surname{Sokalski}}\email{sokalski@el.pcz.czest.pl}
\affiliation{Faculty of Electrical Engineering, Cz\c{e}stochowa University of Technology, Al. Armii Krajowej 17, 42-200 Cz\c{e}stochowa, Poland} \author{Jan \surname{Szczyg{\l}owski}}\email{jszczyg@el.pcz.czest.pl}
\affiliation{Faculty of Electrical Engineering, Cz\c{e}stochowa University of Technology, Al. Armii Krajowej 17, 42-200 Cz\c{e}stochowa, Poland} \affiliation{Eletrotechnical Institute, Ul. Po\.{z}aryskiego 28, 04-703 Warszawa, Poland} \author{Wies{\l}aw \surname{Wilczy\'nski}} \email{w.wilczynski@iel.waw.pl.}\affiliation{Eletrotechnical Institute, Ul. Po\.{z}aryskiego 28, 04-703 Warszawa, Poland}
%\date{}
%\maketitle\\
\begin{abstract}
Data collapse enables comparison of measurement data measured in different laboratories on different samples. In the case of energy losses in Soft Magnetic Materials (SMM) the data collapse is
possible to achieved only if the measurement data can be described by the two components formula. For more complicated cases we propose to perform data collapse's sequence  in the two-dimensional subspaces $L_{i,i+1}$ spanned  by the appropriate powers of frequency $\{f^{i},f^{i+1}\}$. Such approach enables the data comparison in the different two-dimensional subspaces. This idea has been tested with measurement data of the four SMM-s: amorphous alloy  $\textrm{Fe}_{78}\textrm{Si}_{13}\textrm{B}_{9}$,  amorphous alloy
 $\textrm{Co}_{71.5}\textrm{Fe}_{2.5}\textrm{Mn}_{2}\textrm{Mo}_{1}\textrm{Si}_{9}\textrm{B}_{14}$, 
crystalline material -- oriented electrotechnical steel sheets $3\% \textrm{Ni}\textrm{Si}-\textrm{Fe}$,  iron--nickel alloy $79\% \textrm{Ni}\-\textrm{Fe}$.
Intermediate calculations revealed interesting property of the energy losses in the cristalline and amorphous SMM-s which lead to the following hypothesis. Let $P_{tot\,1,2}=f_{1,2}(1+f_{1,2})$ be scaled two-components formula for the energy loss in SMM, where $f_{1,2}$ is the corresponding scaled frequency. Then the scaled energy losses' values in amorphous SMM  are below the second order universal curve $P_{tot\,1,2}=f_{1,2}(1+f_{1,2})$, whereas the scaled energy losses' values in crystalline SMM  are above that universal curve. 
\end{abstract}
\maketitle
 \section{Introduction}\label{I}
 Soft magnetic material (SMM) is a complex system  due to strong nonlinear relation between the magnetic field strength vector
 and the magnetization vector \citep{bib:Fiorillo}. This nonlinearity makes SMM very hard to investigate. In order to achieve progress in theoretical description of the energy loss in SMM a new approach to the theory of the energy loss in SMM has been suggested \citep{bib:Sokal}, \citep{bib:Sokal2}. Therefore, instead of analysis basing on the Maxwell equations \citep{bib:GB},\citep{bib:GB2}  we have assumed that SMM is a complex system which function of energy losses obeys the scaling law. This assumption lead to the total loss energy's formula in a form of general homogenous function:
 \begin{eqnarray}
\exists\hspace{2mm} a,b,c\in { \mathbf{R}}:\hspace{1mm}\label{gen4}\\
 \forall \lambda\in { \mathbf{R}}^{+}\hspace{2mm}  
P_{tot}(\lambda^{a}f,\lambda^{b}B_{m})=\lambda^{c}P_{tot}(f,B_{m}).\nonumber
\end{eqnarray}
Substituting $\lambda=B_{m}^{-\frac{1}{b}}$ we derive general form of $P_{tot}$:
\begin{equation}
\label{general}
P_{tot}(f,B_{m})=B_{m}^{\beta}F\left(\frac{f}{B_{m}^{\alpha}}\right),
\end{equation}
where $F(\cdot)$ is an arbitrary function, $\alpha=\frac{a}{b}$, and $\beta=\frac{c}{b}$. This function depends on features of phenomena to be described. Since our measurement data unable to consider quasi-static losses we choose for $F(\cdot)$ the power series as a rough description of the energy losses  $P_{tot}$ \citep{bib:Sokal}: 
  \begin{eqnarray} 
\label {eq8} 
P_{tot}= B_m^{\beta}\,[\Gamma_{1}\,\frac{f}{B_m^{\alpha}}+ \Gamma_{2}\,\left(\frac{f}{B_m^{\alpha}}\right)^2 +\nonumber\\
\Gamma_{3}\,\left(\frac{f}{B_m^{\alpha}}\right)^3 +\Gamma_{4}\,\left(\frac{f}{B_m^{\alpha}}\right)^4+... ], 
\end{eqnarray}
%\begin{eqnarray}
%\label {eq9} 
%P_{tot}= B_m^{\beta}\,[\Gamma_{1}\frac{f}{B_m^{\alpha}}+ \Gamma_{2}(\frac{f}{B_m^{\alpha}})^2 +... ], 
%\end{eqnarray}
where $f$ - frequency, $B_{m}$ - amplitude of magnetic field's induction. 
Values of $\alpha$, $\beta$ and amplitudes $\Gamma_{n}$  have been estimated for the ten selected soft magnetic materials. It is easy to recognize in (\ref{eq8}) the hysteresis losses $P_{h}$ and the eddy current losses $P_{c}$ by the powers  $f$ and $f^{2}$, respectively. All higher terms correspond to excess losses $P_{ex}$. %{The powers of $B_{m}$ present in the two first terms of (\ref{eq8}) are $\beta-\alpha$ and $\beta-2\,\alpha$. Values of the first exponent calculated with the values  of $\alpha$ and $\beta$ presented in TABLE \ref{Table2} vary from $0.93$ to $1.14$ which is in good agreement with the Bertotti formula for $P_{h}$. However, the second exponent of (\ref{eq8}) corresponding to $P_{c}$ varies from $2.6$ to $3.5$ which overestimates the classical value $2$ by the $30\%-75\%$.   The Bertotti exponents of $B_{m}$ are results of deterministic calculations based on the Maxwell theory. This difference has been interpreted in $\citep{bib:Sokal}$ as an influence of stochastic events inside the sample as well as influence of the  the boudary conditions (sample's shape). Therefore, even if we truncate the  terms higher than the second power the influence of these events are taken account at least partialy.}    
For all samples investigated  in \citep{bib:Sokal},\citep{bib:Sokal2},\citep{bib:yuan},\citep{bib:Sokal3} only the two first terms of (\ref{eq8}) were significant, predicting convex and monotonic increase of $P_{tot}$ v.s. $f$:
\begin{eqnarray}
\label {eq9} 
P_{tot}= B_m^{\beta}\,[\Gamma_{1}\,\frac{f}{B_m^{\alpha}}+ \Gamma_{2}\,\left(\frac{f}{B_m^{\alpha}}\right)^2 ]. 
\end{eqnarray}
Eq.  (\ref{eq9}) has been also confirmed empirically by Yuan et al. \citep{bib:yuan}.
Due to the homogeneity and the second order of (\ref{eq9}), it was possible to obtain the data collapse by appropriate scaling. 
The data collapse consists in a possibility to transform the data of different systems to an universal relation \citep{bib:Gug},\citep{bib:Stan1},\citep{bib:Stan2}. Accordingly, (\ref{eq9}) was transformed to the sample--independent form, which includes the scaled variables $P_{tot\,1,2}$ and $f_{1,2}$:
\begin{equation}
\label{coll}
P_{tot\,1,2} =f_{1,2} +f_{1,2}^2\hspace{2mm},
\end{equation}
where
\begin{equation}
\label{coll2}
P_{tot\,1,2}=\frac{\Gamma_{2}}{\Gamma_{1}^{2}} \frac{P_{tot}}{B_m^{\beta}},
 \hspace{2mm}f_{1,2}=\frac{\Gamma_{2}}{\Gamma_{1}}\frac{f}{B_m^{\alpha}} \nonumber.
\end{equation}
\\$P_{tot\,1,2}$ in (\ref{coll}) we recognize to be a sum of the hysteresis $(P_{h\,1,2})$ and the classical $(P_{c\,1,2})$ losses scaled to dimensionless magnitudes and sample independent representation. The indexes $1,2$ indicate that the corresponding losses are expressed in the $ L_{1,2}$ representation. The aim of this paper is to consider  $P_{tot}$ in different representations, which are the most appropriate for the given kind of energy losses. By this way the considered kind of losses' energy becomes dimensionless and sample independent.\\
 Charts of many measurement data, scaled according to (\ref{coll2}), confirm the data collapse for total energy loss of soft magnetic materials (see Figure 5 in \citep{bib:Sokal}, Figures 5 and 6 in  \citep{bib:Sokal2}. See also Figure 4 in  \citep{bib:yuan} and Figure 2 in  \citep{bib:Sokal3}). 
%Fig. \ref{fig_1} depicts the revealed data collapse on the basis of nine samples.   
% \begin{figure}[!t]
%\centering
%\includegraphics[ width=8cm]{Graph2.eps}
%\caption{Part of a distribution of measurements around theoretical curve according to (\ref{coll})-(\ref{coll2})$\citep{bib:Sokal2}$.}
%\label{fig_1}
%\end{figure}
Why the data collapse is so important? Namely, this leads to reduction of formula describing a phenomenon to an universal form which is sample independent \citep{bib:Gug},\citep{bib:Stan1},\citep{bib:Stan2}. Moreover, recently it has been shown how the data collapse enables comparison between experimental data obtained from measurements on different experimental sets as well as on different samples \citep{bib:Sokal3}.    
However, there are materials for which (\ref{eq9}) is not sufficient and terms of the third and fourth order are important. Unfortunately such an extension unable the data collapse in the sense of \citep{bib:Gug},\citep{bib:Stan1},\citep{bib:Stan2}. This disadvantage can be improved by introducing notion of {\em Partial Data Collapse} (PDC)  which consists in scaling and gauge transformation leading to Data Collapse in different two-dimensional subspaces generated by different powers of $f_{1,2}$. 
%Never the less such data semi-collapse is sufficient to perform data intercomparison and to estimate an uncertainty %measure of experimental set $\citep{bib:Sokal3}$. 
The presented work deals with this improvement. On the PDC's basis we derive the scaling approach to separation of the energy losses  in SMM. The paper is organized in the following way. Section \ref{III} provides  the partial data collapse. The experimental data and the estimations of parameters are presented in Section \ref{II}. Conclusions are given in Section \ref{IV}.
%$\widehat{P}_{tot}$
\section{Partial Data Collapse}\label{III}
In order to approach the PDC concept we consider the two subspaces $L_{1,2}$ and $L_{3,4}$ spanned by $\{f,f^{2}\}$ and $\{f^{3},f^{4}\}$, respectively. 
\subsection{PDC in $L_{1,2}$}
Let us express (\ref{eq8}) by $P_{tot\,1,2}$ and $f_{1,2}$:
\begin{equation} 
\label{gauge1}
P_{tot\,1,2}=f_{1,2}\,(1+f_{1,2})+\chi_{1,2}(f_{1,2},\{\Gamma_{i}\})
 \end{equation}
where, 
\begin{equation}
\label{chi12}
\chi_{1,2}(f_{1,2},\{\Gamma_{i}\})=f_{1,2}^{3}\frac{\Gamma_{1}}{\Gamma_{2}^{2}}\left(\Gamma_{3}+f_{1,2}\,\frac{\Gamma_{1}\Gamma_{4}}{\Gamma_{2}} \right)
\end{equation}
is the gauge function belonging to the subspace $L_{3,4}, $. The first term of the right hand side in (\ref{gauge1}) is just formula (\ref{coll}) for the sum of hysteresis and classical losses. However the second one $\chi_{1,2}(f_{1,2},\{\Gamma_{i}\})$ desribes all 
others contributions to the energy losses interpreted here as excess losses $(P_{ex\,1,2})$. 
 The second term of the right hand side in (\ref{gauge1}) depends on a sample by the set $\{\Gamma_{i}\}$, where $i=1,2,3,4$. Subtracting this term from both sides of (\ref{gauge1}) we derive the following relations:
 \begin{eqnarray}
 \delta{P}_{1,2}(f_{1,2})=P_{tot\,1,2}(f_{1,2})-\chi_{1,2}(f_{1,2},\{\Gamma_{i}\}),\label{gauge2}\\
\delta{P}_{1,2}(f_{1,2})= f_{1,2}\,(1+f_{1,2}),\label{gauge2bis}
 \end{eqnarray}
 where (\ref{gauge2}) transforms the experimental data into the PDC  (\ref{gauge2bis}) (Fig. \ref{Fig.6}). Summarizing this subsection we write down the known formula for the energy losses separation:
 \begin{equation}
 \label{gauge3}
{P}_{tot\,1,2}(f_{1,2})=P_{h\,1,2}+P_{c\,1,2}+P_{ex\,1,2}
\end{equation}
\subsection{PDC in $L_{3,4}$}\label{L34}
PDC in $L_{1,2}$ enabled us to compare the sum of hysteresis and edgy current losses collected from different samples Fig.\ref{Fig.6}. In order to derive  analogous formulae for terms describing all other losses (excess) 
we transform  (\ref{eq8}) to the following form:
\begin{equation}
\label{gauge11}
P_{tot\,3,4}=\psi_{3,4}(f_{3,4},\{\Gamma_{i}\})+f_{3,4}^{3}\,(1+f_{3,4})
 \end{equation}
where
\begin{equation}
\label{gauge4}
P_{tot\,3,4}=\frac{\Gamma_{4}^{3}}{\Gamma_{3}^{4}}\frac{P_{tot}}{B_{m}^{\beta}},\hspace{2mm}
f_{3,4}=\frac{\Gamma_{4}}{\Gamma_{3}}\frac{f}{B_m^{\alpha}}
\end{equation}
and  
\begin{equation}
\label{gauge5}
\psi_{3,4}(f_{3,4},\{\Gamma_{i}\})=f_{3,4}\,\frac{\Gamma_{4}^{2}}{\Gamma_{3}^{3}}\left(\Gamma_{1}+f_{3,4}\,\frac{\Gamma_{2}\Gamma_{3}}{\Gamma_{4}} \right)\nonumber
\end{equation}
(\ref{gauge11}) expresses the sum of hysteresis and  classical $(P_{h,3,4}+P_{c\,3,4}$ as well as excess losses $(P_{ex\,3,4})$ by the dimensionless magnitudes $\psi_{3,4}(f_{3,4})$ and $f_{3,4}^{3}\,(1+f_{3,4})$, respectively. 
Performing gauge transformation on (\ref{gauge11}) with respect to $\psi_{3,4}$, we derive the PDC in $L_{3,4}$:
\begin{eqnarray}
\delta{P}_{3,4}(f_{3,4})= P_{tot\,3,4}-\psi_{3,4}(f_{3,4},\{\Gamma_{i}\})\label{gauge21m}\\
 \delta{P}_{3,4}(f_{3,4})= f_{3,4}^{3}\,(1+f_{3,4}). \label{gauge21}
 \end{eqnarray}
PDC-s (\ref{gauge2}) and (\ref{gauge21}) describe completely the data collapse in systems governed by (\ref{eq8}) and truncated above $n=4$.
The general case for PDC in $L_{j,j+1}$ is presenred in APPENDIX.
\section{Experimental data and parameters's estimations}\label{II}
Measurements of the total energy loss $P_{tot}$ were carried out for the four samples of different classes of soft magnetic materials, which exhibit diverse internal structures and magnetic properties: 
\begin {itemize}
\item {P1 - amorphous alloy  $\textrm{Fe}_{78}\textrm{Si}_{13}\textrm{B}_{9}$,}
\item {P2 - amorphous alloy \\
 $\textrm{Co}_{71.5}\textrm{Fe}_{2.5}\textrm{Mn}_{2}\textrm{Mo}_{1}\textrm{Si}_{9}\textrm{B}_{14}$}
\item {P4 -  crystalline material -- oriented electrotechnical steel sheets 3\% Si--Fe,}  
\item {P7 - iron--nickel alloy $79\% \textrm{Ni}--\textrm{Fe}$,}
\end {itemize}
where P-$s$ are abbreviations for figures and tables. 
The measurements of the total energy losses were carried out as a function of maximum induction $B_m$, at fixed values of frequency $f$. The ranges of induction $B_m$, frequency and total energy losses $P_{tot}$ for each sample are presented in TABLE \ref{Table1}.  Thereby, for each magnetic material the set of curves of total energy losses $P_{tot}$ vs. maximum induction $B_m$ and frequency $f$ was obtained. Next, the energy loss measurements were carried out following to the norm IEC60404--2. During the measurement process the shape factor of secondary voltage was equal to 1.111 $\pm $ 0.5\%. The extended uncertainty of obtained measurements (repeatability of measurements specified with standard deviation) was approximately 1.5\%. 
%\begin{widetext}
\begin{center}
%\begin{group}
%\squeezetable
\begin{table}[!t]
\renewcommand{\arraystretch}{1.3}
\caption{Ranges of measured magnitudes}
\label{Table1}
\centering
\begin{tabular}{|c||c|c|c|}
\hline
 Sample & $f[Hz]$ & $B_{m}[T]$ & $P_{tot}[\frac{W}{kg}]$ \\
\hline\hline
$P{1}$ & 10-400 & 0.101-1.201 &0.001-2.666 \\
$P{2}$ & 10-400 & 0.100-0.999 &0.001-1.755	\\
$P{4}$ & 1-500 & 0.116-1.800&0.000-74.519\\
$P{7}$ & 40-400 & 0.100-0.700 &	0.002-2.427\\
\hline 					
\end{tabular}\\ \vspace{1mm}
\end{table}
%\end{group}
\end{center}
%\end{widetext}
\begin{widetext}
\begin{center}
%\begin{group}
%\squeezetable
\begin{table}[!t]
\renewcommand{\arraystretch}{1.3}
\caption{Scaling exponents and coefficients of (\ref{eq8}) }
\label{Table2}
\centering
\begin{tabular}{|c||c|c|c|c|c|c|}
\hline
 Sample & $\alpha$ & $\beta$ & $\Gamma_{1}[\frac{m^{2}}{s^{2}}T^{\alpha-\beta}]$ & $\Gamma_{2}[\frac{m^{2}}{s}T^{2\alpha-\beta}]$ & $\Gamma_{3}[m^{2}T^{3\alpha-\beta}]$ & $\Gamma_{4}[{m^{2}}{s}\,T^{4\alpha-\beta}]$ \\
\hline\hline
$P{1}$ & -2.3468 &	-1.4072 &	  2.25E-03 & 7.96E-06 & -5.19E-09 &  1.76E-12\\
$P{2}$ & -1.5190 &	-0.3754 &    2.53E-03	& 6.79E-06 & -6.48E-09 &  2.78E-12\\
$P{4}$ & -2.3723&-1.2947&1.80E-02&2.04E-05&7.68E-09&-1.37E-12\\
$P{7}$ & -2.4373&-1.4010&	2.28E-03	&1.05E-05&	3.08E-07	&-8.38E-10\\
\hline 					
\end{tabular}\\ \vspace{1mm}
\end{table}
%\end{group}
\end{center}
\end{widetext}
The parameters' values of (\ref{eq8}) has been estimated for each sample's measurement data by minimization of $\chi^2$ using Simplex method of Nelder and Mead \citep{recipes}, 
see TABLE \ref{Table2}.
 Using these values we will perform the data processing according  to Section \ref{III}. %&&&&&&&&&&&&&&&&&
\subsection{Presentation of measurement data in $L_{1,2}$}\label{L12m}
\begin{figure}[!t]
\centering
\includegraphics[ width=8cm]{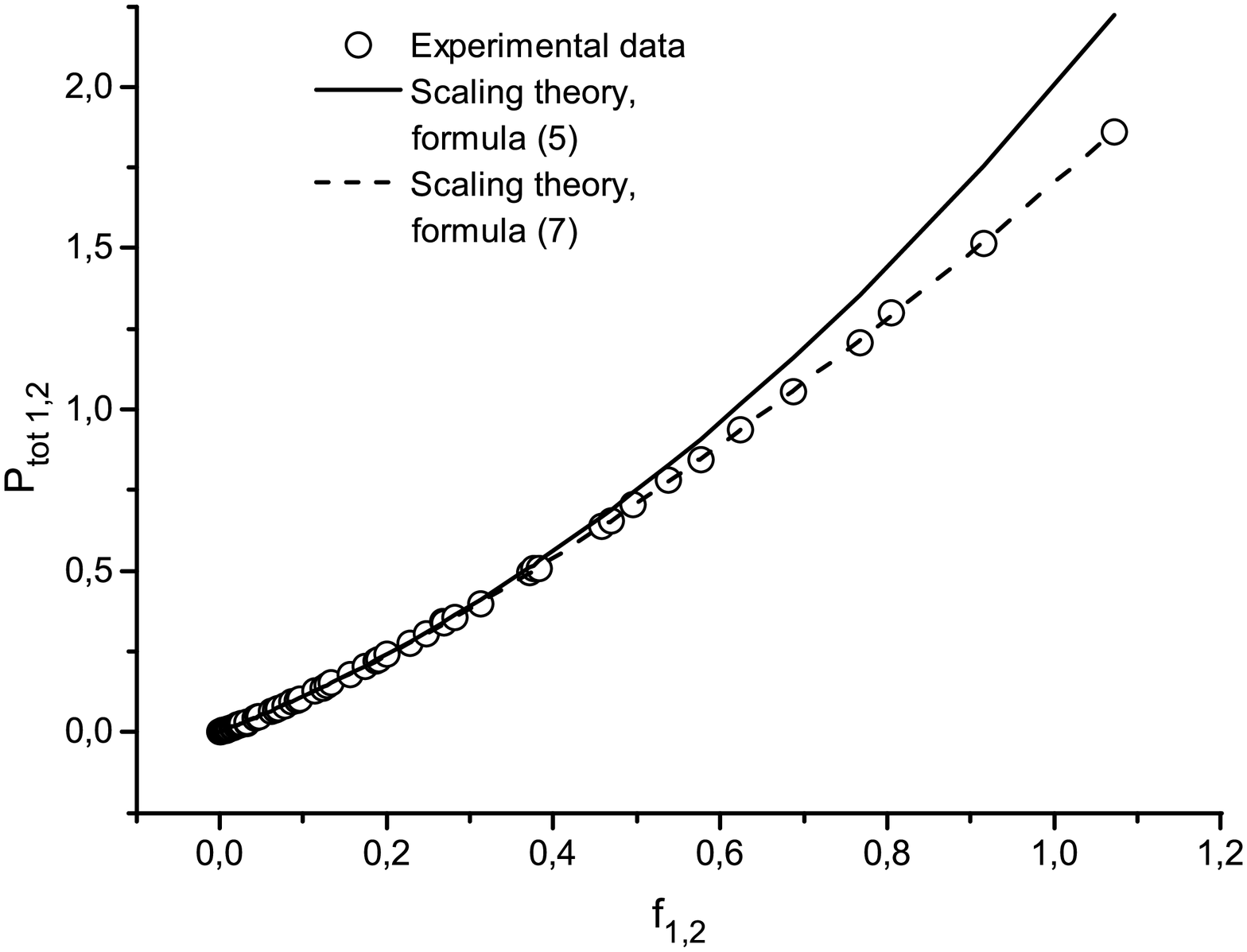}
\caption{$P_{tot\,1,2}$ vs. $f_{1,2}$ for P2 - amorphous alloy 
 $\textrm{Co}_{71.5}\textrm{Fe}_{2.5}\textrm{Mn}_{2}\textrm{Mo}_{1}\textrm{Si}_{9}\textrm{B}_{14}$} 
\label{Fig.2}
\end{figure} 
\begin{figure}[!t]
\centering
\includegraphics[ width=8cm]{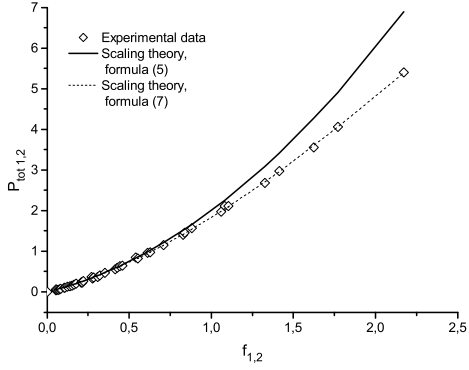}
\caption{$P_{tot\,1,2}$ vs. $f_{1,2}$ for P1 -  amorphous alloy  $\textrm{Fe}_{78}\textrm{Si}_{13}\textrm{B}_{9}$} 
\label{Fig.1}
\end{figure}     
\begin{figure}[!t]
\centering
\includegraphics[ width=8cm]{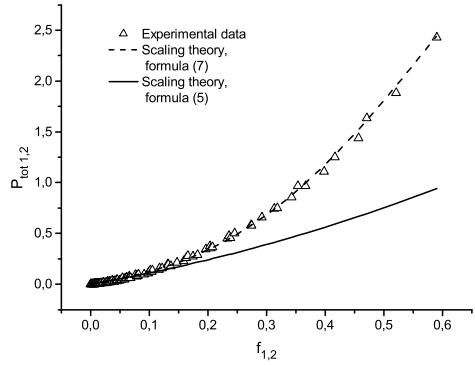}
\caption{$P_{tot\,1,2}$ vs. $f_{1,2}$ for P4 -  crystalline material -- oriented electrotechnical steel sheets $3\% Si--Fe$} 
\label{Fig.3}
\end{figure}   
 \begin{figure}[!t]
\centering
\includegraphics[ width=8cm]{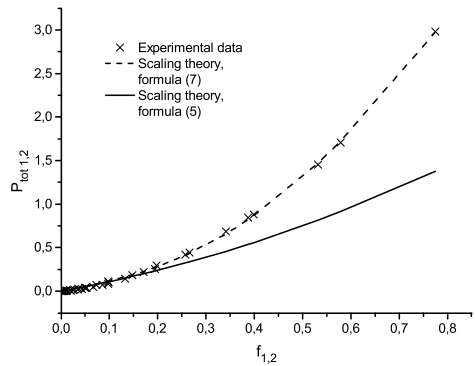}
\caption{$P_{tot\,1,2}$ vs. $f_{1,2}$ for P7 - iron--nickel alloy $79\% \textrm{Ni}--\textrm{Fe}$}
\label{Fig.4}
\end{figure}  
 \begin{figure}[!t]
\centering
\includegraphics[ width=8cm]{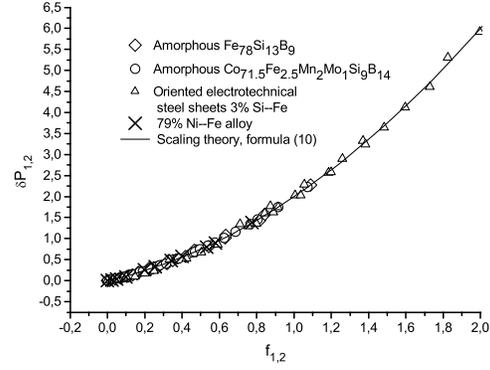}
\caption{Revealed partial data collapse of $\delta{P}_{1,2}$ vs. $f_{1,2}$  }
\label{Fig.6}
\end{figure} 
The obtained results  for the amorphous and crystalline samples were depicted in Fig.\ref{Fig.2}, Fig.\ref{Fig.1} and Fig.\ref{Fig.3}, Fig.\ref{Fig.4}, respectively. Each continuous line presents the universal curve $P_{tot\,1,2}=f_{1,2}(1+f_{1,2})$, see (\ref{gauge2bis}).   The dashed lines present the total energy losses scaled  according to (\ref{gauge1}) and calculated with parameters' values presented in TABLE \ref{Table2}. The markers of points correspond to the  measurement data scaled according  to (\ref{coll2}). For  $f_{1,2}$ which are small enough, the scaled energy losses follows the universal curve (\ref{gauge2bis}). However, above certain value of $f_{1,2}$,  the measurement points and the dashed lines diverge from the universal curve and differences become significant for increasing $f_{1,2}$.  These differences are described by $\chi_{1,2}(f_{1,2},\{\Gamma_{i}\})$, see (\ref{chi12}). Therefore the gauge transformation (\ref{gauge2}) reduces all measurement data and the dashed curves to the common universal curve (\ref{gauge2}). This means that we have achieved  data collapse of the difference between the total energy loss $P_{tot\,1,2}$ and  $\chi_{1,2}$ which is expressed by the square function of $f_{1,2}$ (\ref{gauge2}). 
Plotting $\delta{P}_{1,2}$ vs. $f_{1,2}$ we reveal the data collapse in $L_{1,2}$ for the selected samples Fig.\ref{Fig.6}. We point out that the magnitude which obeys the revealed data collapse is a sum  of hysteresis and classical losses separated from $P_{tot\,1,2}$ (Partial Data Collapse). It is possible to achieve PDC  in any $L_{j,j+1}$ space for which $\Gamma_{j}$ and  $\Gamma_{j+1}$ parameters are relevant (see APPENDIX).\\   
\subsection{Separation curve between cristlline and amorphous phases of SMM }
It is necessary to notice an interesting property of SMM which can be deduced from Fig.\ref{Fig.2} - Fig.\ref{Fig.4} and perhaps can be extended onto  wider sets of SMM. Fig.\ref{Fig.2} and Fig.\ref{Fig.1} suggest that the scaled total energy losses of amorphous SMMs are less than the corresponding values of the universal function (\ref{gauge2bis}) and the scaled total energy losses of crystalline SMMs are greater than the corresponding values of (\ref{gauge2bis}). This behaviour for amorphous materials can be noticed in the papers by Fiorillo \citep{bib:Fiorillo} and Fiorillo et al. \citep{fio2},\citep{fio3} (after scaling of his and of their data). We formulate the following hypothesis: {\em The scaled sum of the hysteresis and classical losses represented by  $\delta{P}_{1,2}(f_{1,2})$ constitutes the phase separator in the plane $( f_{1,2}, P_{tot\,1,2})$ between the crystalline and the amorphous phases of SMM. } 
\begin{figure}[!t]
\centering
\includegraphics[ width=8cm]{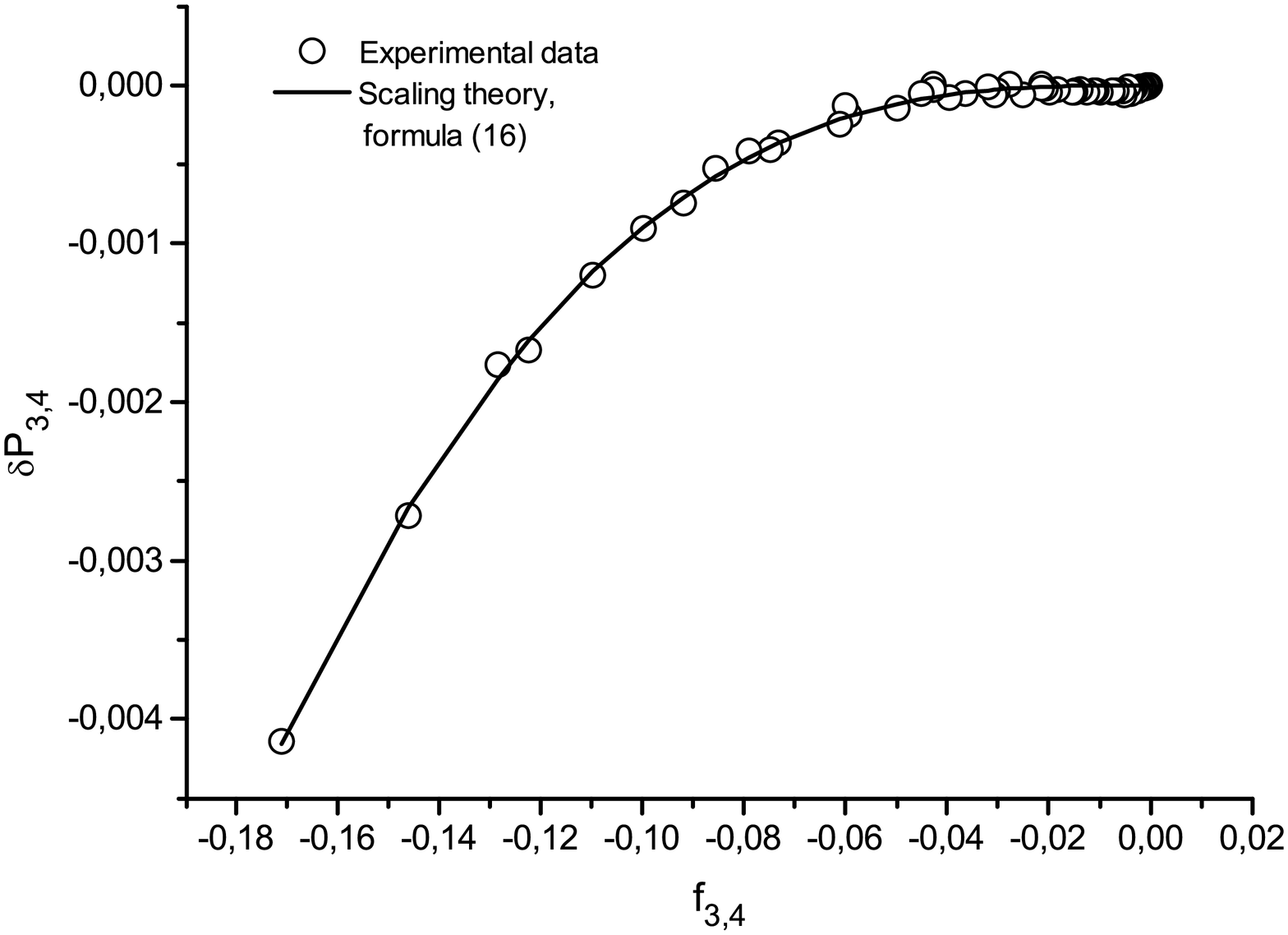}
\caption{$\delta{P}_{3,4}$ vs. $f_{3,4}$ for P2 - amorphous alloy 
 $\textrm{Co}_{71.5}\textrm{Fe}_{2.5}\textrm{Mn}_{2}\textrm{Mo}_{1}\textrm{Si}_{9}\textrm{B}_{14}$} 
\label{Fig.02}
\end{figure} 
\begin{figure}[!t]
\centering
\includegraphics[ width=8cm]{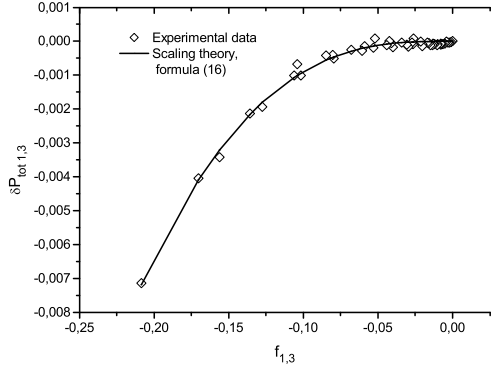}
\caption{$\delta{P}_{3,4}$ vs. $f_{3,4}$ for P1 -  amorphous alloy  $\textrm{Fe}_{78}\textrm{Si}_{13}\textrm{B}_{9}$} 
\label{Fig.01}
\end{figure}     
\begin{figure}[!t]
\centering
\includegraphics[ width=8cm]{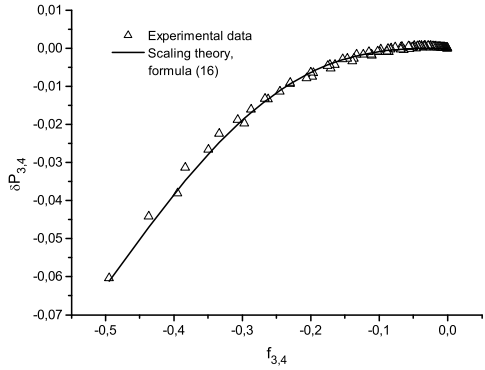}
\caption{$\delta{P}_{3,4}$ vs. $f_{3,4}$ for P4 -  crystalline material -- oriented electrotechnical steel sheets $3\% Si--Fe$} 
\label{Fig.03}
\end{figure}   
 \begin{figure}[!t]
\centering
\includegraphics[ width=8cm]{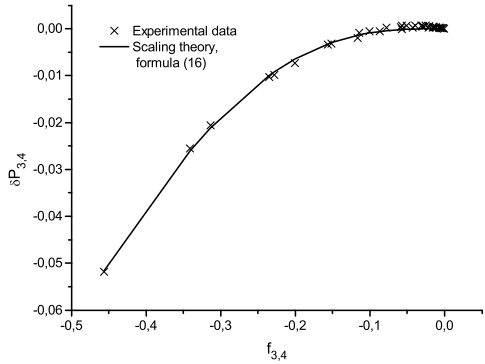}
\caption{$\delta{P}_{3,4}$ vs. $f_{3,4}$ for P7 - iron--nickel alloy $79\% \textrm{Ni}--\textrm{Fe}$}
\label{Fig.04}
\end{figure}  
 \begin{figure}[!t]
\centering
\includegraphics[ width=8cm]{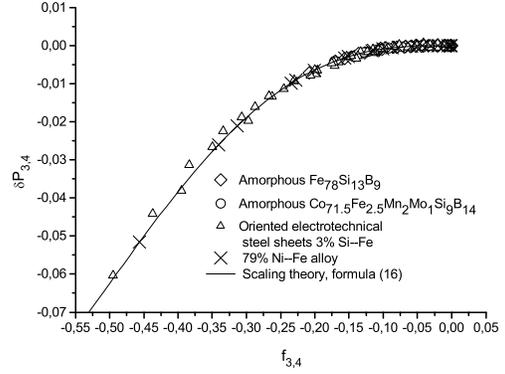}
\caption{Revealed partial data collapse of $\delta{P}_{3,4}$ vs. $f_{3,4}$ }
\label{Fig.06}
\end{figure} 
 \subsection{Presentation of measurement data in $L_{3,4}$}
According to Subsection \ref{L34} we transform the measurement data to the dimensionless and sample independent format which  enables comparison of the excess losses collected from different samples. The obtained results are presented in Fig.\ref{Fig.02} - Fig.\ref{Fig.04}.  The continuous curves are drown according to  (\ref{gauge21}) and represent the dimensional-less and sample independent phenomenological model of the excess energy losses. The points presented by markers correspond to measurement data, scaled according to (\ref{gauge21m}). All these results are presented together in Fig.\ref{Fig.06}, which exhibits PDC of the excess energy losses. It is necessary to explain the negative values of both $f_{3,4}$ and $\delta{P}_{3,4}$. According to (\ref{gauge4}) and TABLE \ref{Table2} for $f>0$ we get $f_{3,4}<0$ for all considered samples. Analogically, using (\ref{gauge21}) and the following constrains: $-0.6<f_{3,4}<0$ (see Fig.\ref{Fig.02}-Fig.\ref{Fig.06}),  we derive the sign of $\delta{P_{3,4}}$ to be negative. Note that this discussion does not concern $L_{1,2}$. In this section we have presented PDC in $L_{1,2}$ and $L_{3,4}$ spaces. 
However it is also possible to transform the measurement data to 
$L_{2,3}$ and get PDC in the form  $f_{2,3}^{2}\,(1+f_{2,3})$. However, we do not find this dependence interesting since $L_{2,3}$ space corresponds to a sum of the classical and a part of the excess losses.
\section{Conclusions}\label{IV}
The assumed scaling reduces the number of independent variables of the energy losses' function from two variables to the effective one (\ref{gen4}), (\ref{general}). We have chosen (\ref{eq8}) rather as generator of a function satisfying  (\ref{gen4}) than an expansion series. 
However, the two first terms suit very well to the common interpretation of the energy losses separation:  $~f$ and $~f^{2}$  hysteresis losses  and classical ones, respectively, whereas the sum of all higher terms can be interpreted as the excess losses. 
The powers of $B_{m}$ appearing in the two first terms of (\ref{eq8}) are $\beta-\alpha$ and $\beta-2\,\alpha$. Values of the first exponent calculated with the values  of $\alpha$ and $\beta$ presented in TABLE \ref{Table2} vary from $0.93$ to $1.14$ which is in good agreement with the Bertotti formula for $P_{h}$. However, the second exponent of (\ref{eq8}) corresponding to $P_{c}$ varies from $2.6$ to $3.5$ which overestimates the classical value $2$ by the $30\%-75\%$.  This difference we explain by the existence of the nonlinear interaction between   the edgy currents and magnetization field \citep{bib:Sokal2}. This interaction has been discussed in many papers, however the derivation of the energy losses formula was done with the linear Maxwell's theory. Our approach is phenomenological, however the general formula (\ref{general}) results from the assumption which states that the considered system is complex one and scale invariant.
By the appropriate scaling we have derived different representations $P_{tot\,j,j+1}$ of $P_{tot}$. Each j-representation basis on the two-dimensional space $L_{j,j+1}$ spanned by the two powers of the scaled frequency: $\{f_{j,j+1}^{j},f_{j,j+1}^{j+1}\}$. Each representation is dimensionless and contains characteristic binomial $f_{j,j+1}^{j}(1+f_{j,j+1})$ which additionally is sample independent. Just this term enables to perform the Partial Data Collapse. 
PDC enables comparison of the energy losses measured on different samples. An idea of energy losses' inter-comparison for the measurement data taken in different laboratories has been published  in \citep{Sievert}. The reason why this comparison was not very successful was the lack of the common measure of error for the energy losses. In the case of measurement data obeying (\ref{eq9}) such a measure has been introduced in \citep{bib:Sokal3}. In this work we have extended notion of this measure for any degree of  (\ref{eq8}). At the end we derive some conclusions concerning the measurement data. The universal curve (\ref{gauge2bis}) is a subspace where the excess losses vanish. Therefore, all points above  this curve correspond to the positive excess energy losses, whereas those below (\ref{gauge2bis}) correspond to the negative values of $P_{ex\,1,2}$. In the light of the revealed separation of energy losses in crystalline and amorphous Soft Magnetic Materials (Subsection \ref{L12m}) we derive the conclusion that $P_{ex\,1,2}^{cristall}\ge 0$ whereas $P_{ex\, 1,2}^{amorphous}\le 0$. Let us note that this conclusion is formulated for the measurement data presentd in representation for $j=1$. The presented PDC is universal method and  can be applied to any experimental data which obey the scaling low.
\section{APPENDIX,  PDC in $L_{j,j+1}$}\label{IIs}%#############################################
Let us assume that (\ref{eq8}) is extended up to the $n$-th order. Then for $j< n$, the corresponding expression for $P_{tot\,j,j+1}$
 reads:
\begin{widetext}
\begin{equation}
\label{gen1}
P_{tot\,j,j+1}=\psi_{j,j+1}(f_{j,j+1},\{\Gamma_{i}\})+f_{j,j+1}^{j}\,(1+f_{j,j+1}) +\chi_{j,j+1}(f_{j,j+1},\{\Gamma_{i}\}),
\end{equation}
\end{widetext}
where $i=1,2,\dots,n$, $j=1\dots, n-1$, 
\begin{equation}
\label{Pf}
P_{tot\,j,j+1}=\frac{\Gamma_{j+1}^{j}}{\Gamma_{j}^{j+1}}\,\frac{P_{tot}}{B_{m}^{\beta}},\hspace{2mm}
f_{j,j+1}=\frac{\Gamma_{j+1}}{\Gamma_{j}}\,\frac{f}{B_{m}^{\alpha}},
\end{equation}
\begin{equation}
\psi_{j,j+1}(f_{j,j+1},\{\Gamma_{i}\})=\sum_{k=1}^{j-1}\Gamma_{k}\frac{\Gamma_{j+1}^{j-k}}{\Gamma_{j}^{j-k+1}}f_{j,j+1}^{k},
\end{equation}
\begin{equation}
\chi_{j,j+1}(f_{j,j+1},\{\Gamma_{i}\})=\sum_{k=2}^{n-j}\Gamma_{j+k}\frac{\Gamma_{j}^{k-1}}{\Gamma_{j+1}^{k}}f_{j,j+1}^{j+k}.
\end{equation}
For analysis of the above magnitudes' dimensions  it is important to know the following dimensions:
\begin{equation}
\label{dim}
\Gamma_{j}\left[m^{2}s^{j-3}T^{\alpha\,j-\beta}\right],
\end{equation}
whereas the following magnitudes are dimensionless: $f_{j,j+1},P_{tot\,j,j+1},\chi_{j,j+1},\psi_{j,j+1},\delta{P}_{j,j+1}$.
Performing the gauge transformation on $P_{tot\,j,j+1}$ we derive the general form of (\ref{gauge2}):
\begin{widetext}
\begin{equation}
\label{gen3}
\delta{P}_{j,j+1}=P_{tot\,j,j+1}-\chi_{j,j+1}(f_{j,j+1},\{\Gamma_{i}\})-\psi_{j,j+1}(f_{j,j+1},\{\Gamma_{i}\})=f_{j,j+1}^{j}\,(1+f_{j,j+1}).
\end{equation}
\end{widetext}
This achievement enable us to formulate the following theorem.\\
Let a phenomenon be described by the function of two independent variables $P_{tot}(f,B)$, let $P_{tot}(f,B)$ be generalized homogenous function (\ref{gen4}). Let the order of source expansion (\ref{eq8}) be $n\ge 2$.
Let the following lists be measurement data governed by the assumed relation $P_{tot}=P_{tot}(f,B)$ :
\begin{equation}
\label{gen5}
[f_{1},\dots,f_{N}],[B_{1},\dots,B_{N}],[P_{tot,1},\dots,P_{tot,N}],
\end{equation}
where $N$ is a number of measured points. Then, there exists a sequence of the $n-1$ scaling+gauge transformations leading to the partial data collapses of (\ref{gen5}), for even (odd) $n$ the number of independent transformations is $\frac{n}{2}$, $(\frac{n+1}{2})$, respectively.
Sections \ref{I} -- \ref{III} constitute the proof of this theorem.  
$\Box$\\ 
There is not need to perform all of them. In order to cover the full space generated by the complete base: $\{f,f^{2},\dots,f^{n}\}$ it is sufficient to transform $(\ref{gen5})$ by the every second transformation of $(\ref{gen3})$.
Summarizing this section we give the interpretation of PDC. First we notice that  for each $j=1,2,...,n-1$  the formula (\ref{gen1}) describes the energy losses in the same sample, whereas $j$ labels different representations of the same formula.  All magnitudes governed by (\ref{gen1}) are dimensionless, however only the following term $f_{j,j+1}^{j}\,(1+f_{j,j+1})$ is sample independent. Due to this term we obtain the dimensionless and sample independent formula (\ref{gen3}). Therefore, the choice of $j$ depends on the terms which  we wish to separate in the dimensionless and sample independent form. For instance,  we have considered the formula (\ref{gen3}) for $n=4$ and $j=1,3$. By this way we obtained PDCs spanned by the following polynomials $f_{1,2}\,(1+f_{1,2})$  and $f_{3,4}^{3}\,(1+f_{3,4})$, respectively.

\bibliographystyle{plainnat}
\end{document}